\documentclass[runningheads]{llncs}

\usepackage{amsmath}
\usepackage{multirow}
\usepackage[table,xcdraw]{xcolor}
\usepackage{graphicx}
\usepackage{textcomp}
\usepackage{subfig}
\usepackage[utf8x]{inputenc}
\usepackage{csquotes}
\usepackage{flexisym}
\newcommand{\ignore}[1]{}
\usepackage[hidelinks]{hyperref}
\usepackage{hyphenat}
\usepackage{wrapfig}

\usepackage[ruled,vlined,linesnumbered]{algorithm2e}

\usepackage{amssymb}
\usepackage[bottom,multiple]{footmisc}
\usepackage{wrapfig}
\usepackage{nicefrac}
\usepackage{bm}
\usepackage{mathrsfs}

\usepackage[font=small]{caption}

\DeclareMathSymbol{\mlq}{\mathord}{operators}{'134}
\DeclareMathSymbol{\mrq}{\mathord}{operators}{'42}

\usepackage{mathtools}

\begin{document}
\title{Practical Aspect of Privacy-Preserving Data Publishing in Process Mining\thanks{\scriptsize{Funded under the Excellence Strategy of the Federal Government and the L{\"a}nder. We also thank the Alexander von Humboldt (AvH) Stiftung for supporting our research.}}}
\titlerunning{Practical Aspect of Privacy-Preserving Data Publishing in Process Mining}
%
%
\author{Majid Rafiei\orcidID{0000-0001-7161-6927} \and
	Wil M.P. van der Aalst\orcidID{0000-0002-0955-6940}}
\authorrunning{Majid Rafiei and Wil M.P. vand der Aalst}
%
\institute{Chair of Process and Data Science, RWTH Aachen University, Aachen, Germany \\
 }
\maketitle              

\begin{abstract}
Process mining techniques such as process discovery and conformance checking provide insights into actual processes by analyzing event data that are widely available in information systems. These data are very valuable, but often contain sensitive information, and process analysts need to balance confidentiality and utility. Privacy issues in process mining are recently receiving more attention from researchers which should be complemented by a tool to integrate the solutions and make them available in the real world. 
In this paper, we introduce a Python-based infrastructure implementing state-of-the-art privacy preservation techniques in process mining. The infrastructure provides a hierarchy of usages from single techniques to the collection of techniques, integrated as web-based tools. 
Our infrastructure manages both standard and non-standard event data resulting from privacy preservation techniques. It also stores explicit privacy metadata to track the modifications applied to protect sensitive data.


\end{abstract}
\section{Introduction}\label{sec:introduction}
Process mining provides fact-based insights into actual business processes using event data, which are often stored in the form of event logs. 
The three basic types of process mining are \textit{process discovery}, \textit{conformance checking}, and \textit{process enhancement} \cite{van2016process}.
An event log is a collection of events, and each event is described by its attributes. The main attributes required for process mining are \textit{case id}, \textit{activity}, \textit{timestamp}, and \textit{resource}.
Some of the event attributes may refer to individuals, e.g., in the health-care context, the \textit{case id} attribute may refer to the patients whose data are recorded, and the \textit{resource} attribute may refer to the employees performing activities for the patients, e.g., nurses or surgeons. 

Privacy issues in process mining are highlighted when the individuals' data are included in the event logs. According to the regulations such as the European General Data Protection Regulation (GDPR) \cite{voss2016european}, organizations are compelled to take the privacy of individuals into account while analyzing their data.
The necessity of responsibly analyzing private data has recently resulted in more attention for privacy issues in process mining \cite{rafieiWA19,rafiei2019role,pretsaICPM2019,MannhardtKBWM19}. 
In \cite{elpaas}, the authors introduce a web-based tool, ELPaaS, implementing the privacy preservation techniques introduced in \cite{pretsaICPM2019} and \cite{MannhardtKBWM19}. ELPaaS gets the required parameters from users and provides results, as CSV files, in email addresses of the users.    

\begin{wrapfigure}{r}{0.480\textwidth} 
	\centering
	\includegraphics[width=0.48\textwidth]{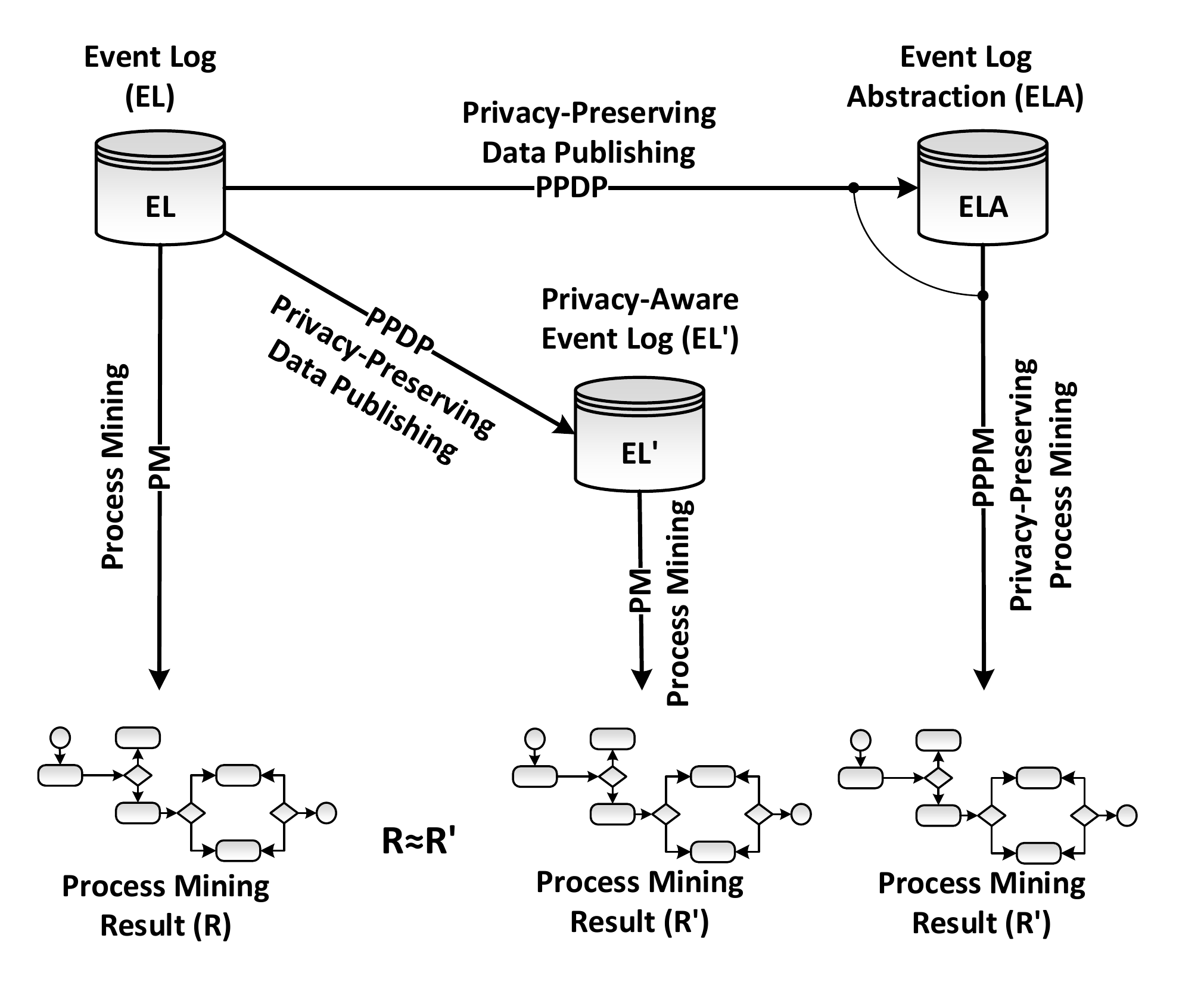}
	\caption{The general approach of privacy in process mining.}\label{fig:general_approach}
	
\end{wrapfigure}

\autoref{fig:general_approach} shows the general approach of privacy in process mining including two main activities: \textit{Privacy-Preserving Data Publishing} (PPDP) and \textit{Privacy-Preserving Process Mining} (PPPM). PPDP aims to hide the identity and the sensitive data of record owners in event data to protect their privacy. PPPM aims to extend traditional process mining algorithms to work with the non-standard data resulting from some PPDP techniques. Note that PPPM algorithms are tightly coupled with the corresponding PPDP techniques.

In this paper, we introduce a tool which mainly focuses on PPDP and offers state-of-the-art privacy preservation techniques including the \textit{connector method} for securely discovering processes \cite{rafieiWA18,rafieiWA19}, the \textit{decomposition method} for privacy-aware role mining \cite{rafiei2019role}, and \textit{$TLKC$-privacy model} for process mining \cite{rafieitlkc}. The \textit{privacy metadata} proposed in \cite{rafieippdp} are also embedded in the offered privacy preservation techniques. Moreover, privacy in the context of process mining is presented through PM4Py-WS (PMTK) \cite{Bertipm4pyWeb} with a web-based interface which is a particular example to show that the provided privacy preservation techniques can be added to the existing process mining tools for supporting PPPM.     

The remainder of the paper is organized as follows. In Section~\ref{sec:demonstration}, we demonstrate the functionality and characteristics of the tool. Section~\ref{sec:maturity} outlines the maturity and availability of the tool, and  Section~\ref{sec:conclusions} concludes the paper.

\section{Functionality and Characteristics}\label{sec:demonstration}
\begin{figure}[bt]
	\centering
	\frame{\includegraphics[width=0.85\textwidth]{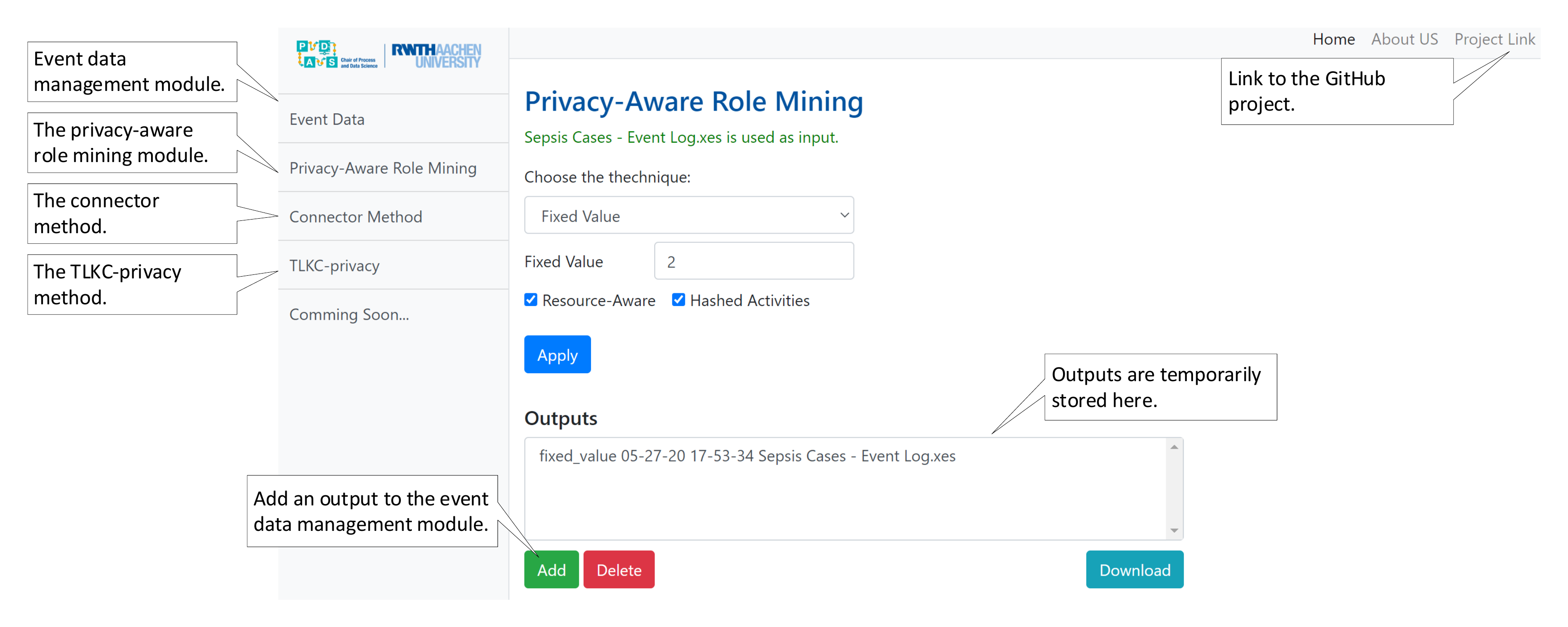}}
	\caption{The privacy-aware role mining page in PPDP-PM.}\label{fig:role_mining}
	\vspace{-5mm}
\end{figure}

In this section, we demonstrate the main functionalities and characteristics of our stand-alone web-based tool, PPDP-PM, which is written in Python using \textit{Django} framework\footnote{\scriptsize https://www.djangoproject.com/}. Our tool has four main modules: \textit{event data management}, \textit{privacy-aware role mining}, \textit{connector method}, and \textit{TLKC-privacy}.   
The \textit{event data management} module has two tabs to upload and manage the event data that could be standard XES event logs\footnote{\scriptsize http://www.xes-standard.org/} or non-standard event data, called \textit{Event Log Abstraction} (ELA) \cite{rafieippdp}. In this module, an event log can be set as the input for the privacy preservation techniques.
The \textit{privacy-aware role mining} module (\autoref{fig:role_mining}) implements the decomposition method supporting three different techniques: \textit{fixed-value}, \textit{selective}, and \textit{frequency-based} \cite{rafiei2019role}. After applying a technique, the privacy-aware event log in the XES format is provided in the corresponding \enquote{Outputs} section. The generated event log preserves the data utility for mining roles from \textit{resources} without exposing who performs what.

\begin{figure}[tb]
	\centering
	\frame{\includegraphics[width=0.80\textwidth]{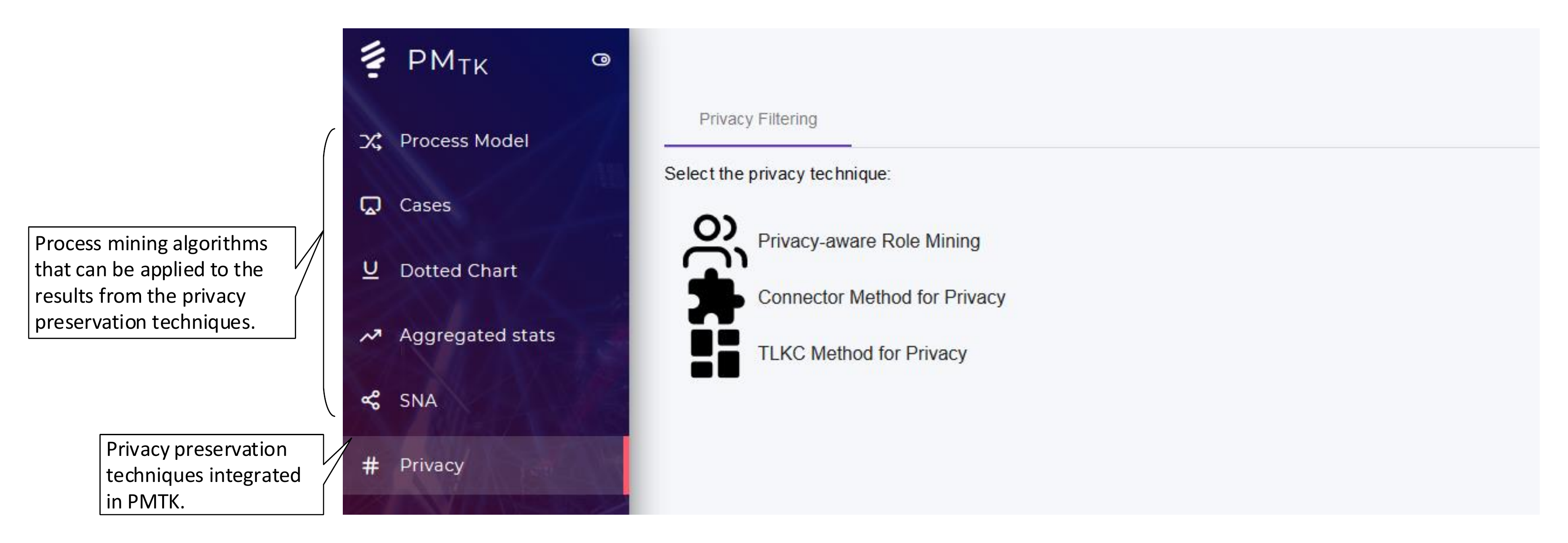}}
	\caption{The home page of the privacy integration in PM4Py-WS (PMTK).}\label{fig:pmtk}
	\vspace{-5mm}
\end{figure}

The \textit{connector method} implements an encryption-based method for discovering directly follows graphs \cite{rafieiWA18,rafieiWA19}. It breaks the traces down into the collection of directly-follows relations which are securely stored in a data structure. After applying the method, the privacy-aware event data are provided in the corresponding \enquote{Outputs} section as an XML file with the ELA format \cite{rafieippdp}.
The \textit{$TLKC$-privacy} module implements the $TLKC$-privacy model for process mining \cite{rafieitlkc} that provides group-based privacy guarantees assuming four types of background knowledge: \textit{set}, \textit{multiset}, \textit{sequence}, and \textit{relative}. $T$ refers to the accuracy of timestamps in the privacy-aware event log, $L$ refers to the power of background knowledge, $K$ refers to the $k$ in the $k$-anonymity definition \cite{sweeney2002k}, and $C$ refers to the bound of confidence regarding the sensitive attribute values in an equivalence class. Applying this method results in a privacy-aware event log in the XES format that preserves data utility for process discovery and performance analysis.
We also provide the same privacy preservation techniques in the context of an open-source process mining tool.
\autoref{fig:pmtk} shows a snippet of the home page of the privacy integration in PMTK where process mining algorithms can directly be applied to the privacy-aware event data.


Each privacy preservation technique in the tool is implemented as a \textit{Django application} that enables the simultaneous running of different techniques on an event log. This architecture makes the whole project easy to maintain, and new techniques can simply be integrated as independent applications. The outputs for the privacy preservation techniques are provided independently for each technique and can be downloaded or stored in the event data repository.
PPDP-PM is designed in a way that provides a cycle of privacy preservation techniques, i.e., the privacy-aware event data, added to the event data repository, can be set as the input for the techniques again as long as they are in the form of standard XES event logs. 
To keep the process analysts aware of the modifications applied to the privacy-aware event logs, the \textit{privacy metadata} \cite{rafieippdp} specify the order of the applied privacy preservation techniques. Moreover, the tool follows a naming approach to uniquely identify the privacy-aware event data based on name of the technique, the creation time, and name of the event log.          


\section{Availability and Maturity}\label{sec:maturity}
As mentioned, PPDP-PM is a web-based application written in Python. The source code, a screencast, and other information are available in a GitHub repository: \url{https://github.com/m4jidRafiei/PPDP-PM}.
The privacy preservation techniques, explained in Section~\ref{sec:demonstration}, and the integration into PMTK are also available as separate GitHub repositories.\footnote{\scriptsize https://github.com/m4jidRafiei/} 
To facilitate the usage and integration of the privacy preservation techniques, they are also published as standard Python packages (https://pypi.org/): \href{https://pypi.org/project/pp-role-mining/}{\textit{{pp-role-mining}}},  \href{https://pypi.org/project/p-connector-dfg/}{\textit{{p-connector-dfg}}}, \href{https://pypi.org/project/p-tlkc-privacy/}{\textit{{p-tlkc-privacy}}}, and \href{https://pypi.org/project/p-privacy-metadata/}{\textit{{p-privacy-metadata}}}.
Our infrastructure provides a hierarchy of usages such that users can use each technique independently, they can use PPDP-PM which integrates a set of privacy preservation techniques as a stand-alone web-based application, and they can also use the provided techniques in a process mining tool where the privacy preservation techniques are integrated.  
The scalability of the tool varies w.r.t. the privacy preservation technique and the size of the input event log. Based on our experiments, our tool can handle real-world event logs, e.g., the BPI challenge datasets\footnote{\scriptsize https://data.4tu.nl/repository/collection:event\_logs\_real}.
However, it can still be improved for industry-scale usage.
PPDP-PM and its integration in PMTK are also provided as Docker containers which can simply be hosted by the users: \url{https://hub.docker.com/u/m4jid}.

\section{Conclusion}\label{sec:conclusions}
Event data often include highly sensitive information that needs to be considered by process analysts w.r.t. the regulations. In this paper, we introduced a Python-based infrastructure for dealing with privacy issues in process mining. A web-based application was introduced implementing privacy-preserving data publishing techniques in process mining. We also showed the privacy integration in PMTK as an open-source web-based process mining tool. 
The infrastructure was designed in such a way that other privacy preservation techniques can be integrated. We plan to cover different perspectives of privacy and confidentiality issues in process mining, and novel techniques are supposed to be integrated into the introduced framework. We also invite other researchers to integrate their solutions as independent applications in the provided framework.

\bibliographystyle{splncs04}
\bibliography{Refrences}

\end{document}